\documentclass[doublecol]{epl2} 
\usepackage{graphicx}
\usepackage{dcolumn}
\usepackage{bm}
\usepackage{color}
\usepackage{amssymb}
\title{SU(3) Landau-Zener Interferometry}
\shorttitle{SU(3) Landau-Zener Interferometry}
\author{M.N. Kiselev\inst{1}, K. Kikoin\inst{2,1}, M.B. Kenmoe\inst{1,3}}
\shortauthor{M.N. Kiselev \etal}
\institute{\inst{1} The Abdus Salam International Centre for Theoretical Physics,
Strada Costiera 11, I-34151 Trieste, Italy\\
\inst{2} School of Physics and Astronomy, Tel-Aviv University, Tel-Aviv
69978, Israel\\
\inst{3} Mesoscopic and Multilayer Structures Laboratory, Faculty of Science, Department of Physics,\\ 
~~ University of Dschang, Cameroon}
\date{\today}
\abstract{
We propose a universal approach to Landau-Zener problem in a three-level system. The problem is formulated in terms of Gell-Mann operators which generate SU(3) algebra and map the Hamiltonian on the effective anisotropic pseudospin 1 model. The vector Bloch equation for the density matrix describing the temporal evolution of three-level crossing problem is also derived and solved analytically for the case where the diabatic states of the SU(3) Hamiltonian form a triangle. This analytic solution is in excellent quantitative agreement with numerical solution of Schr\"odinger equation for a 3-level crossing problem. The model demonstrates oscillation patterns which radically differ from the standard patterns 
for two-level Landau-Zener problem. The triangle works as an interferometer  and the interplay between two paths results in formation of "beats" and "steps"  pattern in the time-dependent transition probability. The characteristic time scales describing the "beats" and "steps" depend on a dwell times in the
triangle. These scales are related to the geometric size of interferometer. The possibilities of experimental realization of this effect in triple quantum dots and in two-well traps for cold gases are
discussed.}
\begin{document}
\maketitle
%\pacs{73.21.La, 
%73.63.Kv,
%73.40.Gk,
%03.75.Lm,
%33.80.Be
%}
{\it 1. Introduction.}  The paradigmatic problem of level crossing known as Landau-Zener  model (LZM) \cite{lan}-\cite{Majorana} is studied for eight decades (see \cite{Nakamura2002} for a review). Various manifestations of LZM are found in all branches of physical sciences from astrophysics to material science.  Recent progress in nanotechnology and cryogenics allow observation and application of  LZM in  quantum transport \cite{quantr}, spintronics \cite{spintr},  nano-magnetism \cite{Sessoli1999}, cold gases, including optical lattices  \cite{Bloch}, mass transport \cite{Trotzky2008}, quantum information processing \cite{Loss98}-\cite{Hanson2007},  etc. 

Standard approach to LZM is based on the universal SU(2) physics of 
two energy levels of the same symmetry which cross by 
linear variation of a control parameter (time, coordinate, energy, chemical potential, 
flux etc). The two states follow a 
diabatic basis or adiabatic (hybridized) basis under fast or slow 
variation of the control parameter. The probability to find a system in a given diabatic/adiabatic state at long time after passing through the crossing point is given by a simple universal one-parametric equation. 
Periodic (non-linear) sweep of control parameter of LZM resulting in 
multiple  passages through the crossing point allows manipulation of interference 
patterns by controlling the St\"uckelberg oscillations associated with the phases accumulated along adiabatic and non-adiabatic paths \cite{Shevchenko2010}. 
%The many-body version of the LZ theory describes creation of the topological defects following a quantum %quench \cite{Quench}. 
The two-level crossing LZ theory is of paramount importance for the theory of  adiabatic quantum computations \cite{Loss98}-\cite{Hanson2007}. Recent progress in nano\-tech\-no\-lo\-gies opened a  new possibility to use LZ interferometry for qubit manipulations  \cite{Galperin2013}. The charge (Josephson)  qubits are manipulated by changing gate voltage (magnetic flux) \cite{Nakamura1997},\cite{Mooji1999}. The spin  qubits are controlled by magnetic field \cite{Loss98}. Technologically, it is more convenient to manipulate  qubits by electric field (gate voltages) \cite{Hayashi2003}. 

\begin{figure}
\begin{center} 
\vspace*{-7mm}
\includegraphics[angle=0,width=\columnwidth]{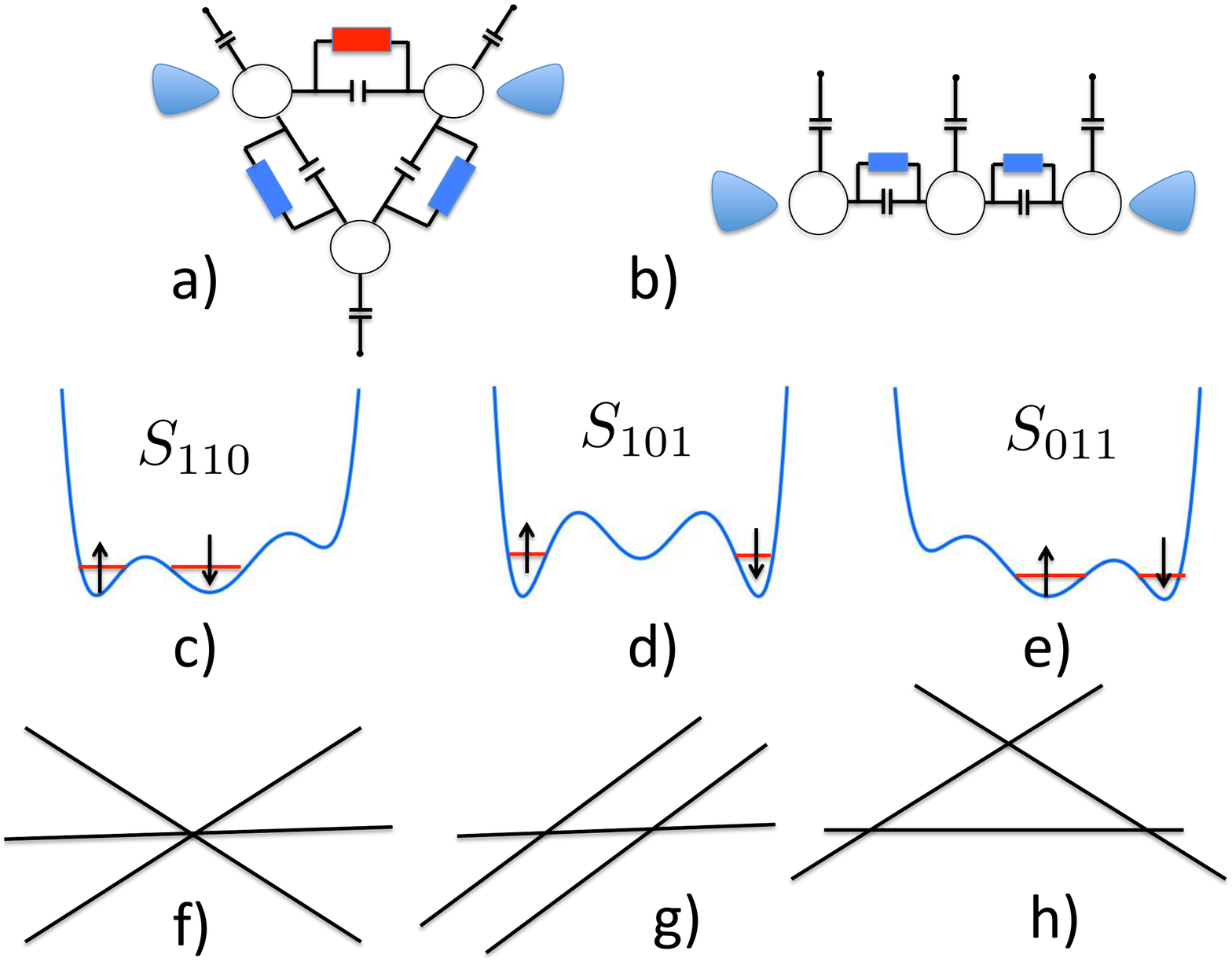}
\vspace*{-10mm}
\end{center}
\caption{(Color online) a) Triangular (TTQD) and b) Linearly arranged triple quantum dot (LTQD). \color{black} Blue and red resistors control the tunneling between dots. Each dot is gated separately. \color{black} c)-e)
 Three singlet states for TQD occupied by two electrons controlled and manipulated by the gate voltage.
f)-h) Three-level crossing described by Hamiltonians (\ref{e1}) - (\ref{e3}).} \label{f.11}
\end{figure}
%\end{center}       
%\vspace*{-8mm}      

In this paper we propose universal tools for description of 3-level  LZM describing qutrits rather than qubits. Instead of Pauli matrices representing SU(2) symmetry of 2-level LZM, we use Gell-Mann matrices forming the basis for SU(3) group describing dynamical symmetry of 3-level systems. We 
formulate generic Hamiltonians for all possible symmetric 3-level configurations and derive the system of Bloch equations describing evolution of density matrix. Numerical solution of Schr\"odinger equation and approximate analytical solution of Bloch equation for the density matrix of 3-level system demonstrate excellent quantitative agreement.  We show that the fingerprint of 3-level LZM is a specific form of  interference oscillations which differ qualitatively from those in 2-level LZM. The shape of these oscillations depends both on the geometrical size of device and on the parameter of adiabaticity.  

{\it 2. Modelling three-level systems.}   
We discuss two possible experimental realizations of three level systems: (i) spinless cold atoms in a
double-well trap (DWT) \cite{Trotzky2008} and 
(ii) triple quantum dots in a triangular (TTQD) \cite{Amaha09} and linearly arranged (LTQD) 
\cite{Schroer07} geometry.      
      
The prototype devices for the 3-level LZM are triple quantum dots (TQD) confining two electrons in a spin singlet states \cite{Petta}-\cite{Van} and double-well traps in optical lattices \cite{Amaha09} confining two spinless cold atoms. The basic features of our model systems are illustrated in Fig. 1. Two possible quantum transport realizations of this regime are triangular \cite{Amaha09} and linearly arranged \cite{Schroer07} (upper panel) TQD occupied by two electrons. The three singlet states are formed by pairs of electrons  $S_{110}$, $S_{011}$ and $S_{101}$  occupying two of three minima (middle panel). Three possible schemes of level crossing are shown in the lower panel.

\color{black}

Three states of doubly occupied DWT correspond to three possible occupations (2,0), (1,1) (0,2) of the left and right wells. Let us fix the reference energy in the middle between the left 
and right levels $\varepsilon_{l,r}$, so that the tunable energy difference is
 $\epsilon_l -\epsilon_r= \epsilon(t)$. Having in mind the analogy 
between the three level system and the S=1 model with uniaxial anisotropy, 
we ascribe the pseudospin projection values $\pm 1, 0$ to the states (2,0), (1,1) (0,2), respectively.
Then, the three crossing levels in LZ problem enumerated with accordance with this agreement are
\begin{eqnarray}\label{eterms}
E_1 &=& E_{2,0}= \epsilon(t)+U,\nonumber\\  
E_0 &=&E_{1,1} = 0,\nonumber\\ 
E_{\bar 1}&=& E_{0,2}= -\epsilon(t)+U,
\end{eqnarray}
where $U$ is the hard-core repulsion energy of two bosons in the same well. Time evolution of these
levels corresponding to a triangular configuration with $\epsilon(t) = vt$ is shown in Fig. \ref{f.11}(h). 
The energy $U$ plays part of the parameter of easy-axis anisotropy.

Three possible configurations of the lowest state of TTQD occupied by two electrons are shown in the
upper panel of Fig. \ref{f.11}. If the wells 1,2,3 are inequivalent, i.e. the 
energy levels $\epsilon_{1}\neq \epsilon_{2} \neq \epsilon_{3}$, then, each two-electron configuration 
is characterized by its own energy $E_{12}\neq E_{23}\neq E_{13}$.  
 The spin state of two electrons is always 
singlet due to the indirect exchange via excited levels \cite{Kikoin2012}. Applying in appropriate
way the gate voltage $V_g(t)$ to corresponding dot, one may realize the level crossing. For example,
changing $\epsilon_1(t)$ one moves the levels  $E_{12}(t)$ and  $E_{13}(t)$ relative $E_{23}$ thus 
realizing LZ regime shown in Fig. \ref{f.11}(g). In case of LTQD geometry we still have three levels 
driven by the gate voltages, but transitions between the states $|1\rangle$ and $|\bar 1\rangle$ are 
strongly suppressed like in the case of real spin 1.  

When considering this system one should remember about existence of higher triplet spin states 
which are not immune to both external
 magnetic field and fluctuations of the Overhauser field associated with the hyperfine interactions.
In principle these states may be involved in LZ transitions, so that the dynamical symmetry of
this system will be described by SO(n) Lie groups \cite{Kikoin2012}. 
We leave this problem for future studies.
\color{black}

{\it 3. General classification of three-levels crossing.} Let us start with construction of 
equivalent spin Hamiltonian for 3-level LZ problem. For this sake we introduce the pseudospin 1 
operator $\vec S$
and associate occupation of three crossing levels with its projections 
$|\bar 1\rangle$, $|0\rangle$, $1\rangle$.  
 The first possibility is crossing of all three diabatic levels  at one point
(Fig 1,f) with effective Hamiltonian 
\begin{eqnarray}
H_1=H_{\rm LZ}= vt S^z + \Delta S^x,
\label{e1}
\end{eqnarray}
where $2\Delta$ is a gap separating lower and upper {\it adiabatic} states and $v$ is the rate 
at which energy changes by external source in the limit $\Delta\to 0$
(we adopt the system of units $\hbar=1$). We refer to this model as SU(2) spin $S=1$ LZ model. 
The properties of this model are well known \color{black} (see \cite{multi} for bow-tie model and \cite{Kenmoe2012}
for S=1 SU(2) model). \color{black} The probability 
to remain in the same diabatic states with $S^z=\pm 1$ is determined by $P_{\rm LZ}=\exp(-\pi\delta/2)$, 
where $\delta=\Delta^2/v$ is the dimensionless LZ parameter.

The second possibility is crossing of three levels  at two points (Fig 1,g) with the  
Hamiltonian 
\begin{eqnarray}
H_2=vt (S^z)^2 +\Delta S^x +h S^z.
\label{e2}
\end{eqnarray}
Here $h$ denotes a tunable level splitting $h =E_{\bar1} - E_{1}$.  The limit $h=0$ corresponds to LZ transition when two-fold degenerate level crosses the non-degenerate state. Note that the model is no
more linear in terms of the generators of the SU(2) group (see \cite{griddegen}).

Below we concentrate on the third possibility where the three levels cross pairwise at three points forming a triangle
(Fig 1,h) with the spin Hamiltonian 
\begin{eqnarray}
H_3= vt S^z +\Delta S^x + D (S^z)^2.
\label{e3}
\end{eqnarray}
The last term stands for a "single-ion" easy axis anisotropy $D$. The Hamiltonian is also non-linear 
in terms of the SU(2) basis. The family of Hamiltonians $H_2$ and $H_3$ can be considered as a 
single-parametric SU(3) deformation of the SU(2) LZ Hamiltonian $H_1$. \color{black} 
%(this model belongs to the class of {\it grid models}, see \cite{grid} for discussion). \color{black} 
Less symmetric LZ level crossing diagrams with different velocities $v_i$ and tunnel rates $\Delta_{ij}$
can be considered as well. In any case the LZ Hamiltonian can be expresses via the generators of SU(3)
algebra. Since the model described by
 $H_3$ is of importance for 
experiments in double- and triple- quantum dots \cite{comm} and contains a basic element for LZ experiments in optical 
lattices, we discuss below its properties leaving discussion of the model $H_2$ 
for \cite{TBP}.

{\it 4. Non-adiabatic transition through a triangular interferometer.} The seeming non-linearity 
of the model $H_3$ is easily removed by representing it in terms of the generators of SU(3) group,  
namely, 8 traceless $3\times 3$ Gell-Mann matrices \cite{Georgi} forming a 
set $\vec \lambda =\{\lambda_\alpha\}, \alpha=1\ldots 8$. 
In this basis the LZ Hamiltonian casts 
a simple {\it linear} form describing interaction of $\vec\Lambda=\vec\lambda/2$ with time-dependent 
magnetic field $\vec B(t)$:
\begin{equation}\label{eq41}
H_{3}=
\vec B(t)\cdot\vec{\Lambda}.
\end{equation}
In order to minimize the number of non-zero components of \color{black} $\vec B(t)$, \color{black} 
it is convenient to use the rotated basis. Two versions of rotated basis are discussed below. In particular, only three combinations of $\lambda$-matrices enter the scalar product
\color{black} \cite{com2} \color{black} in the basis of Gell-Mann matrices adjusted for a linear
TQD \cite{Schroer07}, where direct transitions between the states (110) and (011) are forbidden: 
\begin{eqnarray}\label{mu}
 \vec B^T  &=& \{\Delta,\,vt,\,-D/\sqrt{3} \}, \\
 \vec \Lambda^T &=& \{(\lambda_1+\lambda_6)/\sqrt{2}, (\sqrt{3}\lambda_8 +\lambda_3)/2,\,
 (-\sqrt{3}\lambda_3 +\lambda_8)/2\}. \nonumber
\end{eqnarray}

\begin{center} 
\begin{figure}
\vspace*{-8mm}
\includegraphics[angle=0,width=\columnwidth]{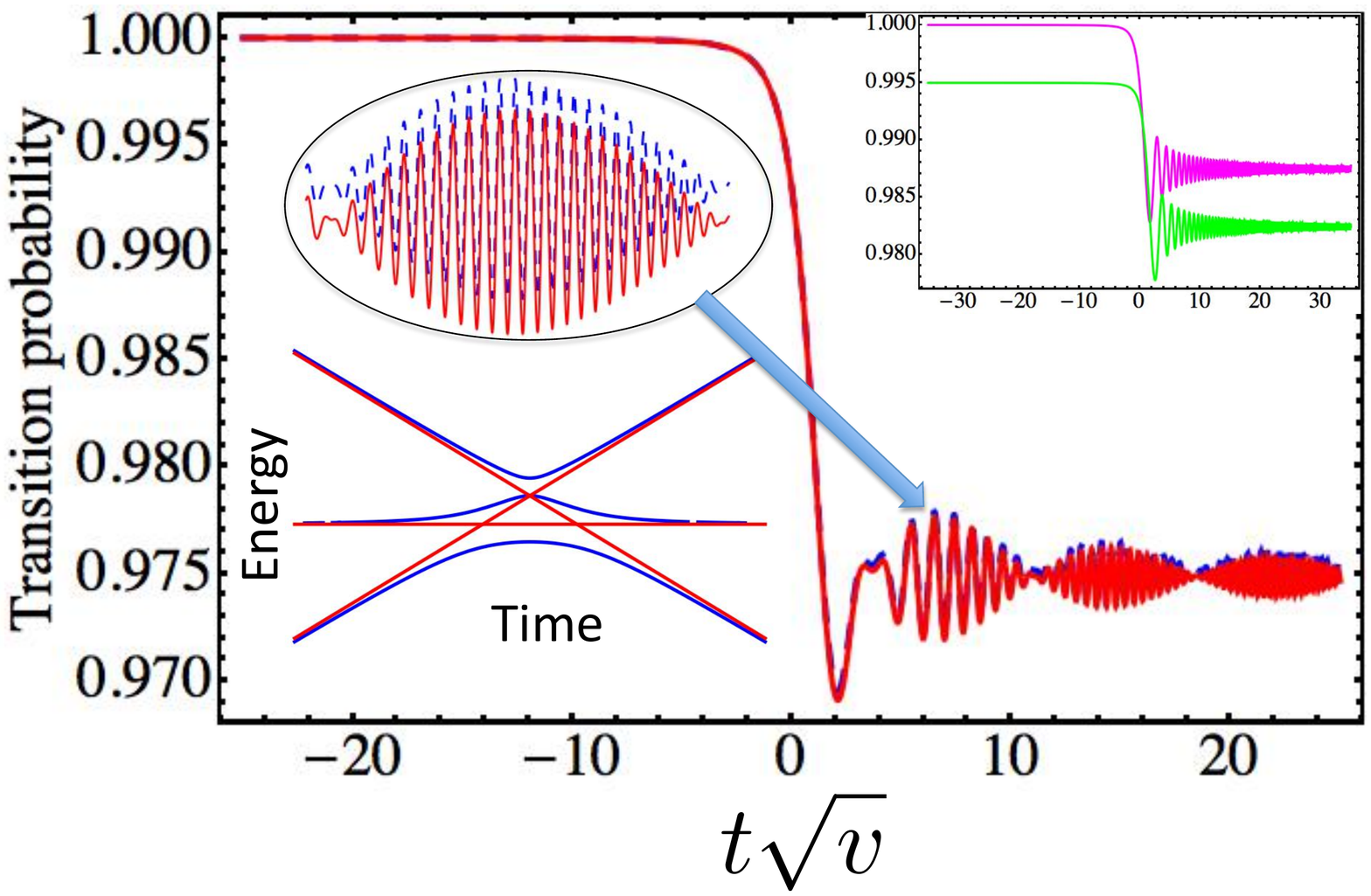}\vspace*{-10mm}
\includegraphics[angle=0,width=\columnwidth]{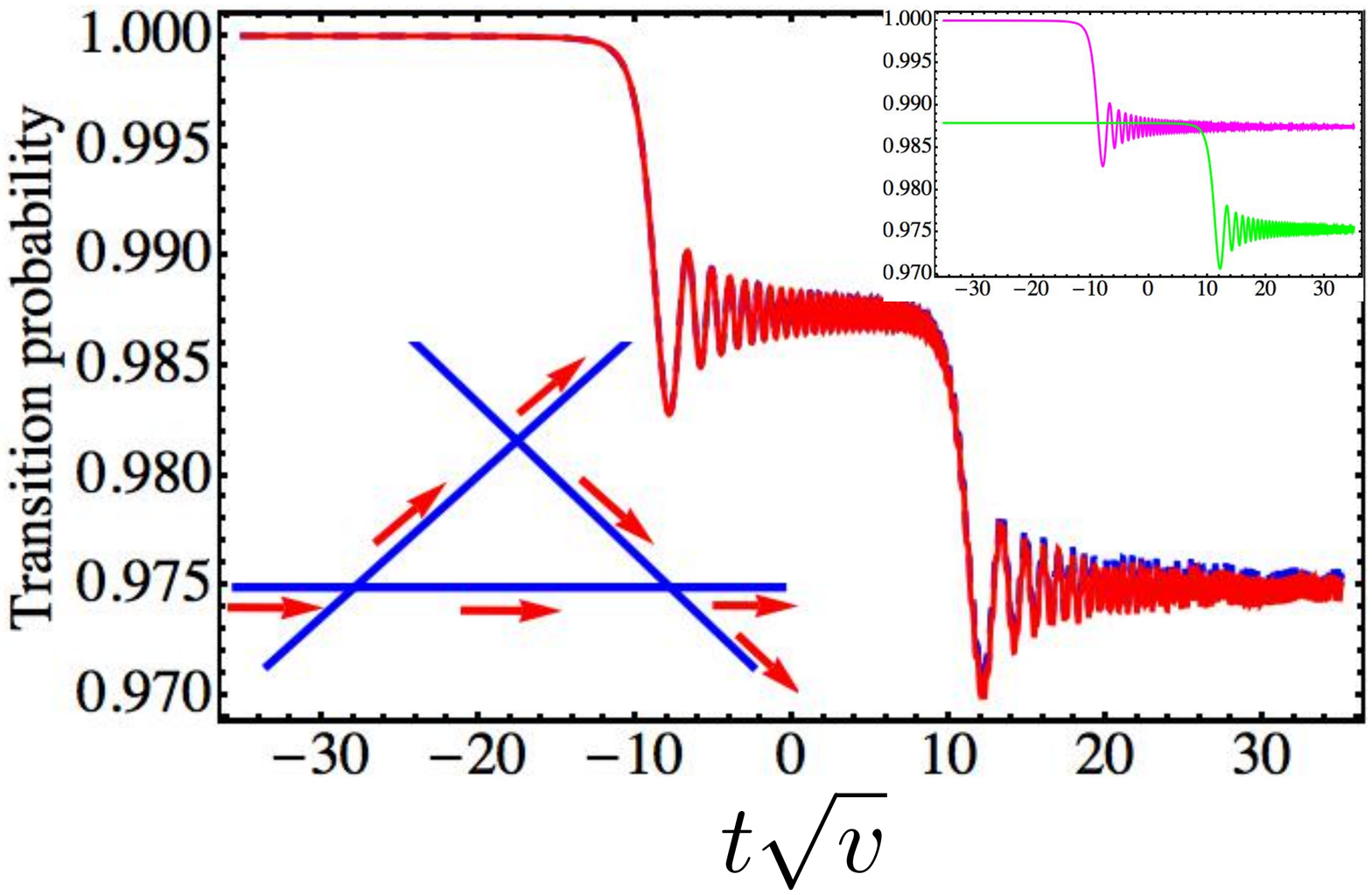}
\vspace*{-12mm}
\caption{(Color online) \color{black} Upper panel: the "beats" {\color{black}calculated at $\Delta^2/v=0.004$, $D/\sqrt{v}=0.425$.} Insert 1: structure of diabatic (red) and adiabatic (blue) states. Insert 2: Zoomed in part of the plot indicated by a blue arrow.
{\color{black}Insert 3: Two level crossing probabilities ignoring the interference term.
The probability
through the first crossing (magenta) at $t_0=0$ is used as the initial condition for the transition through the second crossing (green).} 
Lower panel: the "steps" {\color{black}calculated at $\Delta^2/v=0.004$, $D/\sqrt{v}=10$.} Insert 1: two paths for SU(3) interferometer built out of three singlet states of triple quantum dot. 
{\color{black}Insert 2: Two level crossing probabilities ignoring the interference term. The asymptotic value 
of transition probability through first crossing (magenta) is used as initial condition for the second crossing (green).} 
{\color{black}For both plots} dashed blue line stands for \color{black} numerical \color{black} solution of the Schr\"odinger equation for \color{black} diabatic probability \color{black} $P_{22}$ 
in \color{black} the \color{black} limit $\delta=\Delta^2/v\ll 1$. Solid red line denotes approximate \color{black} analytic \color{black} solution of the Bloch equation (\ref{e5}-\ref{e7}) for $P_{22}$  (\ref{e9}) subject to initial conditions $Q(-\infty)=0, R(-\infty)=-2$ and $W(-\infty)=0$. The solution is valid for arbitrary $\eta=D^2/v$. \color{black} \color{black}
The period of "beats" $T_{beats}\sim 1/D$, the size of "steps" plateau is $T_{steps}\sim D/v$ (see discussion in the text).\color{black}} \label{f.23}
\end{figure}
\end{center}
\vspace*{-10mm} 

The numerical solution for non-adiabatic transition probabilities computed from a 
Schr\"odinger equation with the Hamiltonian $H_3$ is given by a blue dashed curves on Fig. 2. 
Both curves describe a non-adiabatic regime 
$\delta = \Delta^2/v \ll 1$. The left panel demonstrates a "beats" pattern in time dependent 
probability 
when the size of triangle is small $\eta=D^2/v <1$. The right panel shows a "steps" pattern when
the size of triangle is large $\eta=D^2/v >1$. How can we understand these patterns? What are the 
characteristic time scales responsible for this behaviour?

Both "beats" and "steps" are attributed to the interference processes \color{black} 
%(similar behaviour is in fact common for the grid models \cite{chugarin}
%-\cite{bitsteps}).\color{black} 
The triangle formed by  three diabatic states plays a role of LZ interferometer. Schematically, the interference processes
 are shown in the insert 1 of the {\color{black}lower} panel of Fig. 2. The left and upper vertices of the triangle
 work as two splitters while the
right vertex performs mixing. We discuss as an example the transmission probability to remain 
in the same (middle) diabatic state (denoted by $P_{22}$). One possibility to arrive at this state 
is to come along the middle diabatic state. Another one is to go through the upper vertex of the 
triangle responsible also for a "leakage" from the interferometer. The condition whether we get 
"beats" or "steps" should depend on a dwell time in the interferometer. The existence of this
 new pattern is fully attributed to SU(3) symmetry where the dynamics of 
the middle {\it adiabatic} state is non-trivial (see the insert 1 in the {\color{black}upper} panel of Fig. 2), 
being contrasted to trivial dynamics for the symmetric bow-tie model where the diabatic
and adiabatic states are the same.

In order to construct a consistent {\it analytic} description of the SU(3) LZ transition we analyse the equation for the density matrix (DM) in the non-adiabatic limit. The DM can be parametrized by the set of Gell-Mann matrices
\begin{eqnarray}
\hat \rho(t) =\frac{1}{3}\hat{\mathbb{I}} + \frac{1}{\sqrt{3}}\hat{\vec{\lambda}} \cdot \vec n (t),
\label{e4}
\end{eqnarray}
here $\vec n$ is a unit vector in 8-dimensional space of SU(3) generators.
Following a standard procedure we derive a system  of Bloch (von Neumann) equations
\begin{eqnarray}
\frac{d}{d t} \vec n = \vec B(t) \times \vec n(t),
\label{e5}
\end{eqnarray}
where the cross-product is defined as
 $(\vec B\times\vec n)_\alpha = f_{\alpha\beta\gamma} B_\beta n_\gamma$ and $f_{\alpha\beta\gamma}$
 \color{black}
 are totally antisymmetric under exchange of any two indices structure constants of SU(3) group defined by {\it commutation} relations 
$[\lambda_\alpha,\lambda_\beta]=2i f_{\alpha\beta\gamma} \lambda_\gamma$ 
\begin{eqnarray}
f_{\alpha\beta\gamma}=\frac{1}{4i}Tr([\lambda_\alpha,\lambda_\beta]\cdot\lambda_\gamma).
\end{eqnarray}
\color{black}
These equations describe non-dissipative dynamics of the unit vector on a Bloch surface.

\color{black}
In the conventional Gell-Mann basis the generic Hamiltonian $H_{\rm 3}$ 
describing TTQD (Fig.1a) with all three non-zero tunnel matrix elements between dots 
contains five $\lambda$ matrices, 
\begin{eqnarray}
\vec B^T&=&\{vt+D,\, \sqrt{3}vt - D/\sqrt{3},\, \Delta\sqrt{2},\, \Delta\sqrt{2},\Delta\sqrt{2} \},\nonumber\\
\vec \Lambda^T&=&\frac{1}{2}\{\lambda_3,\lambda_8,\lambda_1, \lambda_4, \lambda_6 \}
\end{eqnarray}
Here, the time-dependent level positions $E_{1,2,3}(t)$ are associated with the matrices 
$\lambda_3, \lambda_8$ and inter-level transitions are represented by the matrices 
$\lambda_1, \lambda_4, \lambda_6$. 

The SU(3) LZ Hamiltonian (\ref{eq41}) may be also rewritten in terms of the differences between 
the energy levels (\ref{eterms}), 
by means of appropriate combination of the Gell-Mann matrices \cite{Kikoin2012}. 
Two of three differences, e.g.,
 $E_{10}=\epsilon(t) +U$ and $E_{0\bar 1}=\epsilon(t) - U$ may be chosen.
In this case the effective field $\vec B$ and the spinor $\vec \Lambda$ are  
\begin{eqnarray}\label{rotbas}
 \vec B^T &=& \{2(vt +U)/3,\,2(vt-U)/3,\, \Delta\sqrt{2},\, \Delta\sqrt{2},\,\,\Delta\sqrt{2} \},\nonumber\\
 \vec \Lambda^T &=& \frac{1}{2}\{\lambda_3, \lambda_-,\lambda_1, \lambda_4, \lambda_6\} 
\end{eqnarray}
with $\lambda_{\pm}= (\pm\lambda_3 + \sqrt{3}\lambda_8)/2$.

\begin{figure}[!ht]
\begin{center} 
\vspace*{-5mm}
\includegraphics[angle=0,width=\columnwidth]{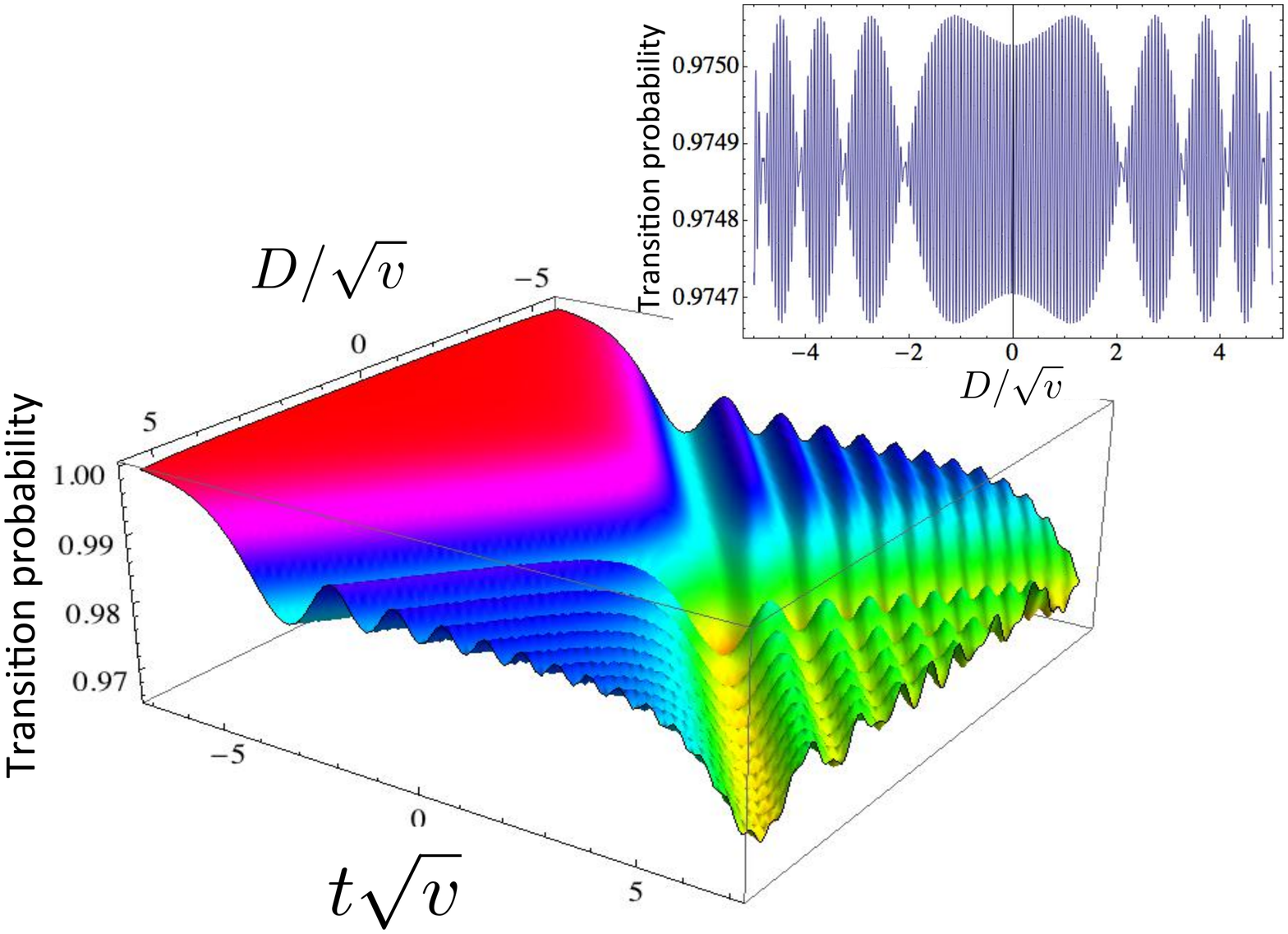}
%\vspace*{-10mm}
%\includegraphics[angle=0,width=\columnwidth]{figure3b.pdf}
\vspace*{-14mm}
\end{center}
\caption{(Color online) \color{black} 
Non-adiabatic transition probability $P_{22}$ as a function of dimensionless time $t/\tau$ and dimensionless uni-axis anisotropy $D\tau$ (LZ time $\tau=1/\sqrt{v}$) computed at $\delta=0.004$. The symmetry $D\to -D$ reflects the symmetry between "easy axis" and "easy plane" anisotropy of zero-dimensional system.
{\color{black}Insert: Transition probability $P_{22}$ at $t\sqrt{v}=100$ as a function of dimensionless dwell
time $\sqrt{v} t_D=D/\sqrt{v}$ computed at $\delta=0.004$ by Eq.\ref{e9}.}
\color{black}} \label{f.33}
\end{figure}
%\vspace*{-1cm}

To minimize the number of components in the model Hamiltonian, we consider the case
of LTQD with suppressed transition $|1\rangle \to |\bar 1 \rangle$, so that the matrix
$\lambda_4$ 
is excluded from $H_{\rm 3}$. This model is straightforwardly mapped
on the S=1 Hamiltonian with easy axis, and one may use a rotated $\mu$-basis of Gell-Mann matrices 
by applying a unitary transformation 
to the $\lambda$-basis
\begin{eqnarray}\label{mugel}
\mu_1&=&(\lambda_1+\lambda_6)/\sqrt{2},\;\;\;
\mu_2=(\lambda_2+\lambda_7)/\sqrt{2},\nonumber\\
\mu_3&=&(\sqrt{3}\lambda_8+\lambda_3)/2,\;\;\;
\mu_4=\lambda_4,\nonumber\\
\mu_5&=&\lambda_5,\;\;\;
\mu_6=-(\lambda_1-\lambda_6)/\sqrt{2},\nonumber\\
\mu_7&=&-(\lambda_2-\lambda_7)/\sqrt{2},\;\;\;\;
\mu_8=(\lambda_8-\sqrt{3}\lambda_3)/2.
\end{eqnarray}
The first three $\mu$-matrices coincide with the SU(2) generators of $S=1$ representation. 
All commutation notations and Casimir operators are preserved. The two-parametric family 
of Landau-Zener Hamiltonians corresponds to one-directional SU(3) deformation of SU(2) LZ model.
The "magnetic field" vector and the spinor $\vec \Lambda$ in this case are
\begin{eqnarray}\label{rotmu}
 \vec B^T  = \{\Delta,\,vt,\,-D/\sqrt{3}\},\;\;\;\;\;\;
 \vec \Lambda^T = \{ \mu_1, \mu_3,\, \mu_8\}. 
\end{eqnarray}
In this representations all interlevel transitions are associated with $\mu_1$, the time-dependent 
level splitting $\epsilon(t)$ is related to $\mu_3$ and the matrix $\mu_8$ is coupled to the anisotropy
parameter.
\bigskip
\color{black}
\begin{widetext}
The eight coupled linear differential equations (\ref{e5}) can be  transferred into a system of three coupled
linear integral equations as follows: 
\color{black}
%\begin{widetext}
\begin{eqnarray}
Q(t)&=&Q(-\infty) -\frac{\Delta^{2}}{2}\int_{-\infty}^t d t_1 \int_{-\infty}^{t_1} d t_2\left(
K_r^{+}(t_1,t_2)Q(t_{2})
+K_r^{-}(t_1,t_2)R(t_{2})\right)
+\frac{\Delta }{2}\int_{-\infty}^t d t_1\Phi_{-}(t_1),\nonumber\\
R(t)&=&R(-\infty) -\frac{3\Delta^{2}}{2}\int_{-\infty}^t d t_1 \int_{-\infty}^{t_1} d t_2\left(
K_r^{+}(t_1,t_2)R(t_{2})
+K_r^{-}(t_1,t_2)Q(t_{2})\right)
+\frac{3\Delta }{2}\int_{-\infty}^t d t_1 \Phi_{+}(t_1),\nonumber\\
W(t)&=&W(-\infty)+\Delta\int_{-\infty}^{t} d t_1\left(K_i^+(t,t_1)
R(t_{1})+ K_i^-(t,t_1)Q(t_{1})\right)  +\Phi_0(t),
\label{e6}
\end{eqnarray}
where
\begin{eqnarray}
\Phi_\pm(t)&=&-\frac{\Delta}{3}\int_{-\infty}^{t}dt_{1}\int_{-\infty}^{t_{1}}dt_{2}\left(\left[
K_r^{20}\cdot K_r^{\pm}-
K_i^{20}\cdot K_i^{\pm}\right]\frac{d}{d t_2} R(t_2)
+\frac{3\Delta}{2}\left[
K_r^{20}\cdot K_i^{\pm}+
K_i^{20}\cdot K_r^{\pm}\right] W(t_2)\right),\nonumber\\
\Phi_0(t)&=&\frac{\Delta}{3}\int_{-\infty}^{t}dt_{1}\int_{-\infty}^{t_{1}}dt_{2}\left(\left[
K_r^{20} \cdot K_i^{+}+
K_i^{20}\cdot K_r^{+}\right]\frac{d}{d t_2}R(t_2)
-\frac{3\Delta}{2}\left[
K_r^{20}\cdot K_r^{+} - 
K_i^{20}\cdot K_i^{+}\right]W(t_2)\right).
\label{e7}
\color{black}
\end{eqnarray}
%\end{widetext}
\color{black}
Here we used the notations $K_{r}^\xi={\rm Re}\exp(i(\xi(t)-\xi(t_1))$, $K_{i}^\xi={\rm Im}\exp(i(\xi(t)-\xi(t_1))$,
$K_{r/i}^\pm=K_{r/i}^{\Omega^{+}}\pm K_{r/i}^{\Omega^{-}}$,
$K_{r/i}^{20}= K_{r/i}^{2\Omega_0}$, and \color{black} $\Omega_0=v t^2/2$\color{black}  , \color{black}   $\Omega^{\pm}=\frac{v}{2}(t\pm\frac{D}{v})^2-\frac{\eta}{2}$\color{black}  .
The product $K_\alpha^\beta \cdot K_\gamma^\delta =K_\alpha^\beta(t_1,t_2)K_\gamma^\delta(t,t_1)$.
The fact that only three real functions are sufficient for complete parametrization of the DM has
 very simple explanation. Let's take the limit $D\to\infty$. In that case we consider three independent 
SU(2) $s=1/2$ LZ transitions accounting also that the tunnel matrix element in the upper vertex of 
triangle scales as $\Delta^2/D$ \cite{com0}. Since the Bloch equations for each of two-level crossings are 
the equations for transition probabilities conserved in each vertex separately, we have just 3 real 
functions to describe this limit.
Since the SU(2) limit $D=0$ is also parametrized by 3 real functions \cite{Kenmoe2012}, 
the upper and lower bound for the number of functions coincide and is equal to 3. 
This reduction is related to some hidden dynamical symmetry connected with higher Casimir invariant of SU(3) group \cite{TBP}.
\end{widetext}

The transition probabilities (diagonal elements of the density matrix) depend only on $Q$ and $R$:
$\rho_{11}=\frac{1}{3}\left(1+\frac{R}{2}+\frac{3Q}{2}\right)$, 
$\rho_{22}=\frac{1}{3}\left(1-R\right)$
$\rho_{33}=\frac{1}{3}\left(1+\frac{R}{2}-\frac{3Q}{2}\right)$. The initial conditions for the lower state occupied at $t=-\infty$ reads:
$Q(-\infty)=R(-\infty)=1$, $W(-\infty)=0$. The initial condition for middle state occupied at $t=-\infty$ is
$Q(-\infty)=0, R(-\infty)=-2$, $W(-\infty)=0$. The initial conditions  for upper state occupied reads: $-Q(-\infty)=R(-\infty)=1$, $W(-\infty)=0$.  We also notice a symmetry $\rho_{11}$$\to$$\rho_{33}$ when 
$Q$$\to$$-$$Q$ which can be used for mapping of LZ transitions with first and third initial conditions \cite{com1}.\\
\bigskip
%\end{widetext}
%\vspace*{5mm}

{\it 5. Results and discussion.} The system of equations can be solved by iterations in the non-adiabatic limit. The analytic solution for
diabatic probability $P_{22}$ in the limit $\delta=\Delta^2/v\ll1$ is given by
\color{black}
\begin{eqnarray}
  P_{22}(t)\approx 1- \pi \delta\left[F\left(t-\frac{D}{v}\right)+F\left(t+\frac{D}{v}\right)\right] +O(\delta^2),
\label{e9}
\end{eqnarray}
\color{black}
where
\begin{eqnarray}
F(t)=\frac{1}{2}\left[\left(\frac{1}{2}+C\left(\sqrt{\frac{v}{\pi}} t\right)\right)^2+\left(\frac{1}{2}+S\left(\sqrt{\frac{v}{\pi}} t\right)\right)^2\right]\nonumber
\end{eqnarray}
and $C(z)$ and $S(z)$ are cosine and sine Fresnel integrals respectively. 
The exact solution can be obtained by exponentiation of the first order expression with correction function
calculated by means of the method elaborated  in \cite{Kenmoe2012}. 
The equation (\ref{e9})  shows two
"waves": one comes from the first splitting at \color{black} $t_-=-D/v$ \color{black} and another one comes from the second
splitter/mixer at \color{black} $t_+=D/v$\color{black}. \color{black} If the period of non-adiabatic oscillations $\tau\sim 1/\sqrt{v}$ \cite{Gefen} is 
{\it large} compared to
a dwell time $t_D\sim D/v$, which is proportional to linear geometric size of the triangle, the two waves interfere constructively forming the "beats" 
pattern (Fig.2 {\color{black}upper} panel). \color{black} In that case 
\begin{eqnarray}
F\left(t-t_+\right)+F\left(t-t_-\right)-2 F(t)\sim \sin(\pi D t) G\left(\sqrt{\frac{v}{\pi}}t\right),
\nonumber
\end{eqnarray}
where $G(z)=\cos(z^2)(S(z)+1/2)-\sin(z^2)(C(z)+1/2)$. \color{black} The period of "beats" is therefore 
$T_{beats}\sim 1/D$. {\color{black} Comparison of these results with the probabilities for two  
independent two-level crossings without interference term (Insert 3 in Fig.2, upper panel) unambiguously demonstrates the key role of the interference processes.} If, however, the period of non-adiabatic 
oscillations is {\it small} compared to the dwell time $\tau<t_D$, 
%the initial state is split twice, the two trains of waves come separately and do not interfere 
the double splitting of initial state (two consequent LZ transitions, {\color{black}see the Insert 2 of the Fig. 2, lower panel}) leads to formation of the "steps" 
pattern (Fig.2. {\color{black}lower} panel). The characteristic time for the "steps" (the size of a plateau) is the dwell time $T_{steps}\sim D/v$. One can see that there are pronounced non-adiabatic oscillations in the plateau of the steps. In order to see both "beats" and "steps" the system should be prepared in any pure state (we showed in Fig.2 and Fig.3 the results corresponding to initial conditions given by occupied middle diabatic state) and the
state $S_{101}$ should be used as a probe for the interference pattern. \color{black} The analytic solution of the Bloch equation (\ref{e6}), (\ref{e7})
 shown by solid red curve demonstrates remarkable agreement with corresponding numerical solution 
of the Schr\"odinger equation (blue dashed curve). The approach based on solution of Bloch equations
 allows one to consider effects of classical fast and slow noise \cite{Kenmoe2012} 
 {\color{black}by either ensemble averaging the {\it  equation} \cite{PokSyn} if the noise is fast or
 averaging its {\it solution} in given realization if the noise is slow \cite{Kenmoe2012}}. 
The noise is associated with fluctuations of the Overhauser field
(double- and triple- quantum dots experiments), fluctuations of electric field (\color{black} Immanuel 
\color{black} Bloch experiments) and fluctuations 
of both charge and flux in superconducting qubits. \color{black} 
Besides, the Bloch equation approach allows treatment of periodically driven LZ systems involving
mixed quantum mechanical states.
We leave these problems for future studies \cite{TBP}. \color{black} 
The results of SU(3) LZ interference are summarized in Fig.3.  The pattern 
%{\color{black}(upper panel)} 
shows the oscillations due to the two-path interference in the non-adiabatic limit.
{\color{black}Transition probability at long times $t\gg t_D$  also shows pronounced "beats" structure
characteristic for two-path interference (see insert of Fig.3).}
% (non-adabaticity parameter $\delta=0.004$ in all calculations). 
The equations (\ref{e5})-(\ref{e7}), the "beats" and "steps" shown in Fig. 2 and the interference 
pattern (Fig.3) represent the central results of this Letter. 

{\it Conclusions.}  We analysed general models of 3-level crossing in the space of SU(3) generators (\ref{eq41}) or pseudospin 1 with anisotropy (\ref{e1})-(\ref{e3}) and presented both numerical
solution of the Schr\"odinger equation
and approximate analytical solution of the Bloch (von Neumann) equation. Excellent agreement between two approaches is demonstrated. If the diabatic states of linearly driven 3-level system form a triangle, it acts as an interferometer with qualitatively new pattern of interference oscillations. Depending on the dwell time in the triangle, the interference pattern shows the "beats" due to constructive interference of two paths or "steps" when two separated in time non-adiabatic LZ transitions take
place. Both "beats" and "steps" are the manifestations of SU(3) symmetry. We believe that the interference pattern predicted in this work can be experimentally observed both in quantum transport (TQD) and in ultra-cold bosons experiments.

We are grateful to B.L. Altshuler, G. Burkard, A. Chudnovskiy, Yu. Galperin, V. Gritsev,  S. Ludwig, G. Mussardo, J.Petta and L. Vandersypen for illuminating discussions. The work of MBK is supported by STEP program at ICTP. MNK appreciates the hospitality at LMU (Munich) where part of this work has been done.
{\color{black}The research of MNK was supported in part by the National Science Foundation under Grant No. NSF PHY11-25915.}


\begin{thebibliography}{99}
\bibitem{com2} Only non-zero components of 8-vector are shown. The constant term $H_0=-\frac{2D}{3}\hat{\mathbb{I}}$ is omitted from the Hamiltonian to shift the middle diabatic state to zero energy.

% Beats and Steps

% Chudnovsky - Garanin
\color{black}


%\bibitem{chugarin} M. F. O'Keeffe, E. M. Chudnovsky, D. A. Garanin, Phys. Rev. {\bf B 87}, 174418 (2013)

%\bibitem{bitsteps} B. M. Garraway and N. V. Vitanov, Phys. Rew. {\bf A 55}, 4418
%(1997); A. Ferrón, D. Domínguez, and M.-J. Sánchez, Phys. Rev. {\bf B 82}, 134522 (2010); 
%K. Saito, M. Wubs, S. Kohler, Y. Kayanuma, and P. Haenggi, Phys. Rev. {\bf B 75},
%214308 (2007).
\color{black}
\bibitem{com1} In the SU(2) limit $D=0$ and $\Phi_-=K^-=0$. The equation for $Q$ represents the vector SU(2) $S=1$ Bloch equation being decoupled from the second and third equations which form a tensor Bloch equation \cite{PokSyn}.
\end{thebibliography}
\end{document}